\documentclass[aps,showpacs,twocolumn]{revtex4}
\usepackage{graphicx}

\begin{document}
%\draft

%<<<<<<<<<<<<< TITLE >>>>>>>>>>>>>>>%
\title{D-braneworld cosmology II: Higher order corrections}

%<<<<<<<<<<<<< AUTHOR >>>>>>>>>>>>>>>%
\author{Tomoko Uesugi$^{(1)}$, Tetsuya Shiromizu$^{(2,3)}$, Takashi Torii$^{(3)}$ and Keitaro Takahashi$^{(4)}$}

%<<<<<<<<<<<<< ADDRESS >>>>>>>>>>>>>>>%

\affiliation{$^{(1)}$Institute of Humanities and Sciences and Department of Physics, 
Ochanomizu University, Tokyo 112-8610, Japan}

\affiliation{$^{(2)}$Department of Physics, Tokyo Institute of Technology, 
Tokyo 152-8551, Japan}

\affiliation{$^{(3)}$Advanced Research Institute for Science and Engineering, 
Waseda University, Tokyo 169-8555, Japan}

\affiliation{$^{(4)}$Department of Physics, The University of Tokyo, 
Tokyo 113-0033, Japan}

%<<<<<<<<<<<<< DATE >>>>>>>>>>>>>>>%
\date{\today}

%======================================%
%<<<<<<<<<<<<< ABSTRACT >>>>>>>>>>>>>>>%
%======================================%
\begin{abstract}
We investigate braneworld cosmology based on the D-brane initiated in our previous paper. 
The brane is described by a Born-Infeld action and the gauge field is contained. The 
higher order corrections of an inverse string tension will be addressed. 
The results obtained by the truncated argument 
are altered by the higher order corrections. The equation of state of the gauge 
field on the brane is radiation-like in low energy scales and almost dust-like fluid in 
high energy scales. Our model is, however, limited below a critical finite value of the energy 
density. For the description of full history of our universe the presence of a S-brane might be essential. 
\end{abstract}


\maketitle
%\vskip1cm

%======================================%
%<<<<<<<<<<<< SECTION I  >>>>>>>>>>>>>>%
%======================================%
%\baselineskip25pt
\label{sec:intro}
\section{Introduction}

Superstring theory is a promising theory to unify interactions. Recent progress 
such as M-theory and discovery of the D-brane implies new picture of the universe. That is, our universe 
is described by a thin domain wall 
in the higher dimensional spacetimes \cite{RSI,RSII,Tess,Roy,cosmos}. Since this 
scenario is motivated by the fundamental feature of the D-brane, it is natural to ask 
what the universe on the D-brane seems to be. We will consider the self-gravitating D-brane because 
we are interested in the effects of high energy. The D-brane is governed by the Born-Infeld 
(BI) action 
when the gauge fields is turned on \cite{BI}. The gauge fields can be regarded as radiation 
on D-brane. 
Hereafter we call the gauge fields on the D-brane BI matter. 

For the self-gravitating D-brane, 
there is a serious issue in supergravity, that is, the BI matter does not play as a source of gravity 
on the brane \cite{DBC3}. This is, however, the case of the zero net cosmological constant. If the
net cosmological constant is non-zero, the BI matter can become a source of gravity \cite{DBC4}. In the 
present paper we consider the model where the bulk stress tensor is described by a negative cosmological 
constant
and the brane follows the BI action with the $U(1)$ gauge field. Bulk fields are turned off. 
In this model, as seen in the next section, the Einstein-Maxwell theory can be recovered at the low 
energy scale. See Ref.~\cite{DBC5,Gas,BIcosmos,BIcosmos2} for other studies on the probe D-brane, 
the brane gas and so on. 

In the BI action, the self-interaction of the gauge field is included in non-linear 
order of the 
%string coupling, 
inverse string tension $\alpha'$. In the previous study \cite{DBC}, we took account of the 
order of $\alpha'^4$ and found the equation of state (EOS) in the homogeneous and isotropic universe. 
The EOS is composed of radiation parts and dark energy parts. Then we used this 
truncated system for the higher energy regime 
to obtain the tendency of the effects of high energy. As a result, it turns out that the BI matter 
behaves as 
radiation at low energy scales and as a cosmological constant at high energy scales. This result has been
applied to the new reheating scenario \cite{DBC2}. 
It should be noted that such truncation is not a
good approximation in principle, although similar arguments are often employed in higher
derivative theories. 

In this paper we will consider all higher order corrections of $\alpha'$ in the homogeneous and 
isotropic universe. The evolution of the universe is clarified beyond the regime where 
the approximation breaks in the previous study. 

The rest of this paper is organized as follows. In Sec.~\ref{sec:model}, we describe our
D-braneworld model.  In Sec.~\ref{sec:cosmo}, we consider the EOS for BI matter. We will see
that the BI matter behaves as a radiation fluid at low energy scales and as a dust-like fluid at
the high
energy regime. Then we study the evolution of the universe on the D-brane. In Sec.~\ref{sec:summery},
we give the summary and the discussion.

%======================================%
%======================================%
%<<<<<<<<<<<< SECTION II  >>>>>>>>>>>>>>%
%======================================%
\label{sec:model}
\section{Model}

In our model, the bulk stress tensor is composed of the negative cosmological constant and 
the brane is described by the BI action;
%===========<Equation>============%
%
\begin{eqnarray}
S_{\rm BI}=-\sigma \int d^4 x {\sqrt {-{\rm det}[g_{\mu\nu}+2\pi\alpha' F_{\mu\nu}]}},
\end{eqnarray}
%
%=================================%
where $F_{\mu\nu}$ is the $U(1)$ gauge field strength. Thus, photon is already included. In
this setting the gravitational equation on the brane is written by \cite{Tess}
%===========<Equation>============%
%
\begin{eqnarray}
{}^{(4)}G_{\mu\nu}=8\pi G T_{\mu\nu}+ \kappa^4 \pi_{\mu\nu}-E_{\mu\nu},\label{Einstein}
\end{eqnarray}
%
%=================================%
where 
%===========<Equation>============%
%
\begin{eqnarray}
8\pi G = \frac{{\kappa}^2}{\ell},\label{scale}
\end{eqnarray}
%
%=================================%
%===========<Equation>============%
%
\begin{equation}
\pi_{\mu\nu} = -\frac{1}{4}T_{\mu\alpha}T^{\;\alpha}_{\nu} +\frac{1}{12}TT_{\mu\nu} 
+\frac{1}{8}g_{\mu\nu}T^{\alpha}_{\;\beta} T^{\beta}_{\;\alpha}-\frac{1}{24}g_{\mu\nu}T^2,
\end{equation}
%
%=================================%
and
%===========<Equation>============%
%
\begin{eqnarray}
E_{\mu\nu}=C_{\mu\alpha\nu\beta}n^\alpha n^\beta.
\end{eqnarray}
%
%=================================%
In the above we have supposed the Randall-Sundrum fine-tuning;
%===========<Equation>============%
%
\begin{eqnarray}
\frac{1}{\ell}=\frac{\kappa^2}{6}\sigma,
\end{eqnarray}
%
%=================================%
where $\ell$ is the curvature length of the five dimensional anti-de Sitter spacetime. 

In four dimensions $S_{\rm BI}$ becomes 
%===========<Equation>============%
%
\begin{eqnarray}
S_{\rm BI} & = & -\sigma \int d^4 x {\sqrt {-g}} \Biggl[
1-\frac{1}{2}(2\pi\alpha')^2{\rm Tr}(F^2) \nonumber \\
 & & ~ +\frac{1}{8}(2\pi\alpha')^4 \bigl({\rm Tr}(F^2)\bigr)^2-\frac{1}{4}(2\pi\alpha')^4 {\rm
Tr}(F^4) \Biggr]^{1/2} .
\nonumber 
\\
\end{eqnarray}
%
%=================================%

Expanding the above action with respect to $\alpha'$, it becomes 
%===========<Equation>============%
%
\begin{equation}
S_{\rm BI}  =  -\sigma \int d^4x {\sqrt {-g}} \Biggl[ 1-\frac{1}{4}(2\pi\alpha')^2{\rm Tr}(F^2) 
+O(\alpha'^4)\Biggr]. 
\label{BIaction}
\end{equation}
%
%
%=================================%

The energy-momentum tensor on the brane is given by
%===========<Equation>============%
%
\begin{eqnarray}
{T}^{\rm (BI)}_{\mu\nu}& = & -\sigma g_{\mu\nu}+ \sigma 
(2\pi \alpha')^2 {T}^{\rm (em)}_{\mu\nu} +O(\alpha'^4),
\label{em-tensor}
\end{eqnarray}
%
%=================================%
where 
%===========<Equation>============%
%
\begin{eqnarray}
{T}^{\rm (em)}_{\mu\nu}= F_\mu^{~\alpha} F_{\nu\alpha}
-\frac{1}{4}g_{\mu\nu}F_{\alpha\beta}F^{\alpha\beta}.
\end{eqnarray}
%
%=================================%
To regard the above $ {T}^{\rm (em)}_{\mu\nu}$ as the energy-momentum tensor of 
the usual Maxwell field on the brane, we set 
%===========<Equation>============%
%
\begin{eqnarray}
\sigma (2\pi\alpha')^2=1.
\end{eqnarray}
%
%=================================%
Substituting Eq.~(\ref{em-tensor}) into Eq.~(\ref{Einstein}), we obtain Einstein-Maxwell 
theory
%===========<Equation>============%
%
\begin{eqnarray}
G_{\mu\nu} \simeq 8\pi G T_{\mu\nu}^{\rm (em)}
\end{eqnarray} 
%
%=================================%
at the leading order. In the next section, we will take higher order corrections into account.

%======================================%
%<<<<<<<<<<<< SECTION III  >>>>>>>>>>>>>>%
%======================================%
\label{sec:cosmo}
\section{Higher order correction to homogeneous and isotropic universe}

\subsection{Equation of state for BI matter}

Let us focus on the homogeneous and isotropic universe.
We consider the single brane model. Then the metric on the brane is 
%===========<Equation>============%
%
\begin{eqnarray}
ds^2=-dt^2+a^2(t) \gamma_{ij}dx^i dx^j.
\end{eqnarray}
%
%=================================%
The modified Friedmann equation becomes \cite{cosmos} 
%===========<Equation>============%
%
\begin{eqnarray}
\label{friedman}
\biggl( \frac{\dot a}{a} \biggr)^2= \frac{\kappa^2}{3 \ell}\rho_{\rm BI}+
\frac{\kappa^4}{36}\rho_{\rm BI}^2-\frac{K}{a^2},
\end{eqnarray}
%
%=================================%
and
%===========<Equation>============%
%
\begin{equation}
\label{raychud}
\frac{\ddot a}{a}=-\frac{\kappa^2}{6\ell}(\rho_{\rm BI}+3P_{\rm BI})-\frac{\kappa^4}{36}
\rho_{\rm BI}(2\rho_{\rm BI}+3P_{\rm BI}).
\end{equation}
%
%=================================%
As long as we consider the homogeneous and isotropic universe, we can omit the contribution from 
$E_{\mu\nu}$ \cite{Tess}. The universe is described by the domain wall in the anti-de Sitter spacetime. 

Let us define the electric and magnetic fields as usual
%===========<Equation>============%
%
\begin{eqnarray}
E^i=F_0^{~i},~~~{\rm and}~~~B^i=\frac{1}{2}\epsilon^{ijk}F_{jk}.
\end{eqnarray}
%
%=================================%
Here we assume that the $E_i$ and $B_i$ are randomly oriented fields and those 
coherent length is much shorter than the cosmological horizon 
scales. Then $\langle E_i E_j \rangle = (1/3)g_{ij}E^2$, 
$\langle B_i B_j \rangle = (1/3)g_{ij}B^2$, $\langle E_i \rangle = \langle B_i \rangle =0$ 
and $\langle E_i B_j \rangle =0$, which are natural in the homogeneous and isotropic universe. 
In addition, it is natural to assume ``equipartition"
%===========<Equation>============%
%
\begin{eqnarray}
E^2(t)=B^2(t)=:\epsilon.
\end{eqnarray}
%
%=================================%
They can be regarded as a part of background radiation. From the above assumption we immediately obtain
the following useful formulae 
%===========<Equation>============%
%
\begin{eqnarray}
\langle {\rm tr}(F^2) \rangle = 2 \langle (E^2-B^2) \rangle =0, 
\end{eqnarray}
%
%=================================%
%===========<Equation>============%
%
\begin{eqnarray}
\langle {\rm tr}(F^4) \rangle & = &  \langle 2(E^2-B^2)^2 \rangle + 4 \langle (E \cdot B)^2 \rangle 
\nonumber \\
& = & \frac{4}{3}E^2 B^2= \frac{4}{3}\epsilon^2 \neq 0,
\end{eqnarray}
%
%=================================%
%===========<Equation>============%
%
\begin{eqnarray}
\langle (F^2)_{00} \rangle =-E^2=-\epsilon,
\end{eqnarray}
%
%=================================%
%===========<Equation>============%
%
\begin{eqnarray}
\langle (F^2)_{i0} \rangle =\epsilon^k_{~ji} \langle E_j B_k \rangle =0 ,
\end{eqnarray}
%
%=================================%
%===========<Equation>============%
%
\begin{eqnarray}
\langle (F^2)_{ij} \rangle & = & \langle ( E_i E_j +B_i B_j -g_{ij}B^2) \rangle 
\nonumber \\
& = &  \frac{1}{3}g_{ij}(E^2+B^2)-g_{ij}B^2= -\frac{1}{3}g_{ij}\epsilon,
\end{eqnarray}
%
%=================================%
%===========<Equation>============%
%
\begin{eqnarray}
\langle (F^4)_{00}\rangle = \langle E^2(B^2-E^2) \rangle - \langle (E \cdot B)^2 \rangle 
=-\frac{1}{3}\epsilon^2,
\end{eqnarray}
%
%=================================%
%===========<Equation>============%
%
\begin{eqnarray}
\langle (F^4)_{0i} \rangle = \langle (E^2-B^2)(F^2)_{0i} \rangle =0,
\end{eqnarray}
%
%=================================%
and
%===========<Equation>============%
%
\begin{eqnarray}
\langle (F^4)_{ij} \rangle  & = & \langle (E_i E_j +B_i B_j -g_{ij}B^2)(E^2-B^2) \rangle 
\nonumber \\
& & ~~+g_{ij} \langle (E \cdot B)^2 \rangle \nonumber \\
& = & \frac{1}{3}g_{ij}\epsilon^2. 
\end{eqnarray}
%
%=================================%

To close the equations we need the EOS of the background radiation. To do this, 
we must evaluate the averaged energy-momentum tensor which can be derived by the action. If we 
keep only the terms which can contribute to the averaged energy-momentum tensor,  
%===========<Equation>============%
%
\begin{widetext}
\begin{eqnarray}
S_{\rm BI} & \supset & - \sigma \int d^4x{\sqrt {-g}}
\Biggl\{1-\frac{1}{4\sigma}{\rm tr}(F^2)-\frac{1}{8\sigma^2}{\rm tr}(F^4) 
-\sum_{n=2}^\infty \frac{(2n-3)!!}{n!}\biggl[ \frac{{\rm tr}(F^4)}{8\sigma^2}\biggr]^{n}
\nonumber \\ & & -2 \sum_{n=2}^\infty \frac{(2n-3)!!}{(n-1)!} \frac{{\rm tr}(F^2)}{\sigma} 
\biggl[ \frac{{\rm tr}(F^4)}{8\sigma^2}\biggr]^{n-1}
\Biggr\}.
\end{eqnarray}
%
%=================================%
Thus, the averaged energy-momentum tensor becomes 
%===========<Equation>============%
%
\begin{eqnarray}
\langle {T}^{\rm (BI)}_{\mu\nu} \rangle 
& = & -\sigma g_{\mu\nu}+\langle {T}^{\rm (em)}_{\mu\nu} \rangle \nonumber \\
& & -\frac{1}{\sigma} \left\langle (F^4)_{\mu\nu}-\frac{1}{8}g_{\mu\nu}{\rm tr}(F^4)  
\right\rangle
-4 \sigma \sum_{n=2}^\infty \frac{(2n-3)!!}{4(n-1)!\;8^{n-1}\sigma^{2n-1}} 
\left\langle (F^2)_{\mu\nu} \bigl[ {\rm tr}(F^4) \bigr]^{n-1}\right\rangle 
\nonumber \\
& & -8 \sigma \sum_{n=2}^\infty \frac{(2n-3)!!}{n!\;8^n \sigma^{2n}} 
\left\langle \Bigl[n(F^4)_{\mu\nu}-\frac{1}{8}g_{\mu\nu}{\rm tr}(F^4) \Bigr]
\bigl[ {\rm tr}(F^4) \bigr]^{n-1}   \right\rangle \nonumber \\
& =: & -\sigma g_{\mu\nu}+\langle T_{\mu\nu} \rangle. 
\end{eqnarray}
%
%=================================%
Then the density and pressure are 
%===========<Equation>============%
%
\begin{eqnarray}
\rho_{\rm BI}  = \langle T_{00} \rangle 
& =  & 
\langle E^2 \rangle +\frac{1}{2\sigma} \left\langle (E \cdot B ) ^2 \right\rangle
+ 4\sigma \sum_{n=2}^\infty \frac{(2n-3)!!}{(n-1)!\;2^{n+1}\sigma^{2n-1}}
\left\langle E^2 (E \cdot B)^{2n-2} \right\rangle 
\nonumber \\
& & + 4\sigma \sum_{n=2}^\infty \frac{(2n-1)!!}{n!\; 2^{n+2}\sigma^{2n}}
\left\langle (E \cdot B)^{2n} \right\rangle  
\nonumber \\
& = & \epsilon + \frac{1}{6}\sigma \Bigl( \frac{\epsilon}{\sigma} \Bigr)^2
+ 4 \sigma \sum_{n=2}^\infty \frac{(2n-3)!!}{(2n-1)(n-1)!\;2^{n+1}} \Bigl(
\frac{\epsilon}{\sigma} \Bigr)^{2n-1} + 4 \sigma \sum_{n=2}^\infty
\frac{(2n-1)!!}{(2n+1)n!\;2^{n+2}} \Bigl( \frac{\epsilon}{\sigma} \Bigr)^{2n}
\nonumber \\
&=& \sigma \biggl[ -1 + \Bigl( \frac{\sigma}{\epsilon} + 1\Bigr) {\rm arcsin}
\Bigl( \frac{\epsilon}{\sigma} \Bigr) \biggr],
\label{rr}
\end{eqnarray}
%
%
%=================================%
and
%===========<Equation>============%
%
\begin{eqnarray}
P_{\rm BI}=\frac{1}{3}\langle T^i_i \rangle 
& = &  
\frac{1}{3}E^2 -\frac{1}{2\sigma}\left\langle (E \cdot B ) ^2 \right\rangle
+\frac{1}{3}4 \sigma \sum_{n=2}^\infty \frac{(2n-3)!!}{(n-1)!\;2^{n+1}\sigma^{2n-1}}
\left\langle E^2 (E \cdot B)^{2n-2} \right\rangle 
\nonumber \\
& & -4\sigma \sum_{n=2}^{\infty} \frac{(2n-1)!!}{n!\; 2^{n+2}\sigma^{2n}}
\left\langle (E \cdot B)^{2n} \right\rangle  
\nonumber \\
& = & \frac{1}{3}\epsilon - \frac{1}{6}\sigma \Bigl( \frac{\epsilon}{\sigma} \Bigr)^2
+ \frac{4}{3} \sigma \sum_{n=2}^\infty \frac{(2n-3)!!}{(2n-1)(n-1)!\;2^{n+1}} \Bigl(
\frac{\epsilon}{\sigma} \Bigr)^{2n-1} - 4 \sigma \sum_{n=2}^\infty
\frac{(2n-1)!!}{(2n+1)n!\;2^{n+2}} \Bigl( \frac{\epsilon}{\sigma} \Bigr)^{2n}
\nonumber \\
& = & \sigma \biggl[1+\Bigl(-\frac{\sigma}{\epsilon}+\frac{1}{3} \Bigr){\rm arcsin}
\Bigl(
\frac{\epsilon}{\sigma} \Bigr)
\biggr],
\label{pp}
\end{eqnarray}
\end{widetext}
%
%=================================%
respectively. In the last lines of Eqs (\ref{rr}) and (\ref{pp}), the radius of the convergence 
is $\epsilon \leq 1$. In the above derivation, we used 
%===========<Equation>============%
%
\begin{eqnarray}
\langle E_{i_1}E_{i_2} \cdots E_{i_{2n}} \rangle = \frac{1}{(2n+1)!!} (\delta_{i_1 i_2}\cdots 
\delta_{i_{2n-1} i_{2n}} + \cdots )\epsilon^{n},
 \nonumber 
\end{eqnarray}
%
%=================================%
and so on. 

\begin{figure}[tbp]
\includegraphics[width=.85\linewidth]{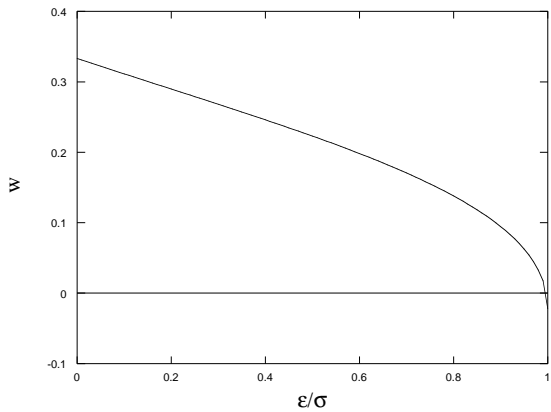}
\caption{
$w$-$\epsilon /\sigma$ diagram of the BI matter.
}
\label{fig:w}
\end{figure}

It is easy to see 
%===========<Equation>============%
%
\begin{equation}
w:=\frac{P_{\rm BI}}{\rho_{\rm BI}} \approx 
\left\{ \begin{array}{ll}
\frac{1}{3} - \frac{2}{9} \frac{\epsilon}{\sigma}, & \mbox{for}\  \frac{\epsilon}{\sigma} \ll 1 \\
-\frac{\pi-3}{3(\pi-1)} + \frac{4 \sqrt{2}}{3 (\pi-1)^{2}} \sqrt{1- \frac{\epsilon}{\sigma}}, & \mbox{for}\
\frac{\epsilon}{\sigma} \approx 1.
\end{array} \right.
\end{equation}
%
%=================================%
Then the BI matter behaves like  radiation and almost dust-like fluid. This means that the universe
cannot be accelerate by BI matter. See Fig.~\ref{fig:w} for the profile of $w$ as a function of
$\epsilon/\sigma$.  This is contrasted with the previous result where we used the truncated
action, that is, we  considered the corrections up to $O(\alpha'^4)$ and investigated the
situation where the  correction terms dominate the lowest order terms. The previous treatment
is similar to  that in a higher derivative theory. In superstring theory, the Einstein equation 
can be derived at the lowest order of $\alpha'$. The higher order correction 
terms  is expressed by the higher derivative terms in general. 

%------------<fig.1>---------------------------
%--------------<fig.1>-----------------------

Before studying the evolution of the universe, we point out an interesting feature. 
As $\epsilon /\sigma \ll 1$ we can expand $\rho$ and $P$ as
%===========<Equation>============%
%
\begin{eqnarray}
\rho= \sum_{n=1}^\infty {}^{(n)}\!\rho ,
\end{eqnarray}
%
%=================================%
%===========<Equation>============%
%
\begin{eqnarray}
P=\sum_{n=1}^\infty {}^{(n)}\!P,
\end{eqnarray}
%
%=================================%
where $ {}^{(n)}\!\rho =O(\epsilon^n)$ and ${}^{(n)}\!P=O(\epsilon^n)$. Then 
we can see 
%===========<Equation>============%
%
\begin{eqnarray}
{}^{(n)}P=\frac{1}{3}{}^{(n)} \rho,~~{\rm for}~~n={\rm odd},
\end{eqnarray}
%
%=================================%
and
%===========<Equation>============%
%
\begin{eqnarray}
{}^{(n)}\rho = -{}^{(n)}P,~~{\rm for}~~n={\rm even}. 
\end{eqnarray}
%
%=================================%
The $n=$odd and $n=$even order parts behave like radiation and vacuum energy, respectively.

%%=================================%
\subsection{Evolution of universe}

By using the energy conservation law, $\dot \rho_{\rm BI}+3H(\rho_{\rm BI}+P_{\rm BI})=0$, 
on the brane, the scale factor dependence of the energy density is fixed as $\rho \sim a^{-3(1+w)}$. At
the low and high energy regimes, $\rho_{\rm BI} \propto a^{-4}$ and $\propto a^{-2\pi/(\pi-1)}$,
respectively. 

To see the qualitative evolution of the universe, the following rearrangement of the generalized Friedmann
equation is useful,
%===========<Equation>============%
%
\begin{eqnarray}
a'^2+V(a)=-\bar{K},
\label{frie}
\end{eqnarray}
%
%=================================%
where 
%===========<Equation>============%
%
\begin{eqnarray}
V(a)=-a^2 \bar{\rho}_{\rm BI} (2+\bar{\rho}_{\rm BI} ),
\end{eqnarray}
%
%=================================%
and we have normalized the variables as $\bar{t}:=t/\ell$, 
$\bar{\rho}_{\rm BI}:=\rho_{\rm BI}/\sigma$, 
$\bar{P}_{\rm BI}:=P_{\rm BI}/\sigma$, $\bar{\epsilon}:=\epsilon/\sigma$,
$\bar{K}:=K \ell^2$. A prime denotes  differentiation with respect to $\bar{t}$.

\begin{figure}[tb]
\includegraphics[width=.85\linewidth]{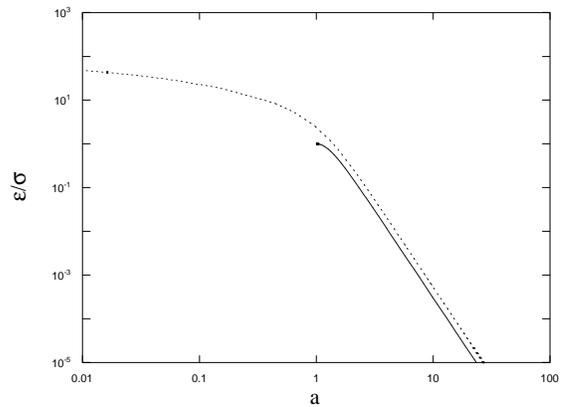}
\caption{
The $a$-$\bar{\epsilon}$ diagram of the full order (solid line) 
and the truncated (dashed line) systems.
}
\label{fig:a-ep}
\end{figure}

{}From the energy conservation law, we find
%===========<Equation>============%
%
\begin{equation}
\bar{\epsilon}'=\frac{4a'}{a}{\rm arcsin}(\bar{\epsilon})
\left[\frac{1}{\bar{\epsilon}^2}{\rm arcsin}(\bar{\epsilon})
-\left( \frac{1}{\bar{\epsilon}}+1\right)
\frac{1}{\sqrt{1-\bar{\epsilon}^2}} \right]^{-1}.
\label{evep}
\end{equation}
%
%=================================%
Solving the Eqs.~(\ref{frie}) and (\ref{evep}) simultaneously, 
we obtain the evolution of the universe.
Fig.~\ref{fig:a-ep} is the $a$-$\bar{\epsilon}$ diagram. For the truncated
system $\bar{\epsilon}$ behaves as $\propto a^{-4}$ in the late time while
it gradually increases beyond $\bar{\epsilon}=1$ as the scale factor becomes small.
This regime is beyond the application of the analysis of the perturbation. On the other
hand, the solution of full order is terminated when $\bar{\epsilon}=1$.
Around $\bar{\epsilon}=1$, the Eq.~(\ref{evep}) is approximated as
%===========<Equation>============%
%
\begin{eqnarray}
\bar{\epsilon}' \simeq -\frac{\pi a'}{a}\sqrt{1-\bar{\epsilon}^2}.
\end{eqnarray}
%
%=================================%
Then we find the solution
%===========<Equation>============%
%
\begin{eqnarray}
\bar{\epsilon}=\cos\left[\pi \ln \left(\frac{a}{a_0}\right)\right].
\end{eqnarray}
%
%=================================%
Hence for the finite value $a=a_0$, $\bar\epsilon$ becomes the critical value $\bar\epsilon=1$.
It is interesting that the physical variables such as $\rho_{\rm BI}$, $P_{\rm BI}$, $\epsilon$
and curvatures do not diverge at this point.

%------------<fig.2>---------------------------
%--------------<fig.2>-----------------------

%------------<fig.3>---------------------------
\begin{figure}[htb]
\includegraphics[width=.85\linewidth]{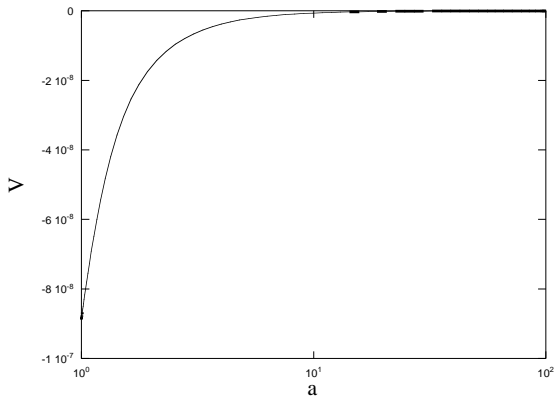}
\caption{
The configuration of the potential function of the generalized Friedmann equation.
}
\label{fig:a-v}
\end{figure}
%--------------<fig.3>-----------------------

%------------<fig.4>---------------------------
\begin{figure}[tb]
\includegraphics[width=.85\linewidth]{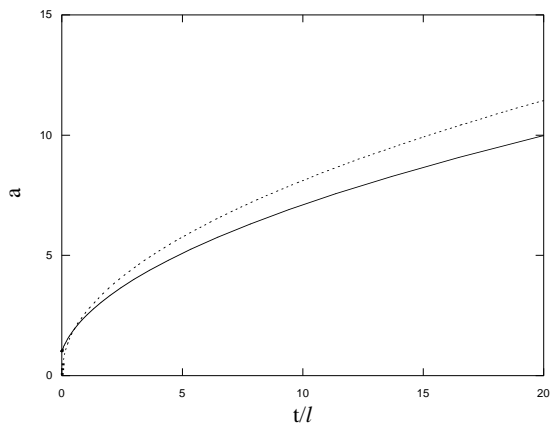}
\caption{
The evolution of the scale factor in the $K=0$ case of the full order (solid line) 
and the truncated (dashed line) systems.
}
\label{fig:ev-t}
\end{figure}
%--------------<fig.4>-----------------------

The configuration of the potential is shown in Fig.~\ref{fig:a-v}. 
In the limit of $\bar{\epsilon}\to 1$ the potential function does not diverge unlike the
ordinary radiation or matter field nor vanish unlike the truncated system in the previous work
but converges non-zero finite value.
We show the evolution of the scale factor in the case of $K=0$ in Fig.~\ref{fig:ev-t}. While in the truncated
system the scale factor diverges with the initial singularity, it starts from finite
value at $t=t_0$ in the full order case.

Then there arises a question: what is the state of the universe before the time $t=t_0$.
If we consider the quantum creation of the universe in this system, we can
expect that the Euclidean action takes minimum value at $a_0$ somehow and that
it becomes very large or even diverges beyond this point. As a result, the 
universe starts from $a=a_0$. 
%However, in this scenario, there is no
%inflation to solve several cosmological problems. Hence the simplest way is to
%add an inflation field and the behavior of the early stage is determined by this
%inflation field.
Other possibility is that there is a reflection symmetry with respect to $t_0$.
Since the scale factor is not connected smoothly, we can guess that there is a S-brane \cite{sbrane} like
structure at $t=t_0$. In this scenario, the universe experiences a bounce. 
However, we have to tune the tension of the S-brane.

It is sure that the evolution of the universe in the early
stage is affected by taking into account the higher order corrections of the
$\alpha'$ expansion.

%======================================%
%<<<<<<<<<<<< SECTION IV  >>>>>>>>>>>>>>%
%======================================%
\section{Summary}
\label{sec:summery}

In this paper we considered the homogeneous and isotropic universe on the D-brane. 
The matter (gauge field) is automatically included in the BI action and 
there are higher order correction terms. The EOS is like 
radiation at the low energy scale and almost dust-like at the high energy scale. In high 
energy scales of braneworld, the $\rho^2$ term  dominates and then the scale factor 
becomes $a(t) \propto t^{(\pi-1)/(2\pi)}$. This model is limited below $\bar \epsilon = 1$ 
similar to the result in the Born and Infeld's original paper \cite{OBI}. 
At the critical value of $\bar \epsilon =1$, however, physical quantities are finite. 
Hence one might want to extend to the past. On a classical level, a mild 
singularity which is a jump of the expansion rate of the universe occurs. To resolve this 
mild singularity an introduction of a S-brane seems to be important.

The present result is quite different from the previous study \cite{DBC} where truncated 
theory was employed. Since the truncation is often used in higher derivative theory 
for non-adequate regime, the previous trial study is still worthy as a first step. From 
the present study, however, we learned that such rough truncation approach does 
not give us good predictions.  

Finally we should comment on our model. We assumed that the bulk stress tensor is 
composed of a negative cosmological constant and the brane action is BI one. 
In Ref. \cite{DBC3}, it is claimed that 
the gauge field cannot be a source of gravity on the D-brane without a cosmological 
constant. To recover the ordinary Einstein equation, the presence of a net 
cosmological constant is essential. We should take into account the above issues to 
obtain a firm picture of D-braneworld cosmology.

%======================================%
%<<<<<<<<< Acknowledgments  >>>>>>>>>>>%
%======================================%
%\baselineskip25pt
\section*{Acknowledgments}

We would like to thank Misao Sasaki and Akio Sugamoto for fruitful discussions. 
To complete this work, the
discussion during and after the YITP workshops YITP-W-01-15 and  YITP-W-02-19
were useful. The work of TS was supported by Grant-in-Aid for Scientific
Research from Ministry of Education, Science, Sports and Culture of 
Japan(No.13135208, No.14740155 and No.14102004). The work of KT was supported by JSPS.

\end{document}